\documentclass[a4paper,twocolumn,10pt]{article}

\usepackage{arxiv}
\usepackage[utf8]{inputenc} % allow utf-8 input
\usepackage[T1]{fontenc}    % use 8-bit T1 fonts
\usepackage{hyperref}       % hyperlinks
\usepackage{url}            % simple URL typesetting
\usepackage{booktabs}       % professional-quality tables
\usepackage{amsfonts}       % blackboard math symbols
\usepackage{nicefrac}       % compact symbols for 1/2, etc.
\usepackage{microtype}      % microtypography
\usepackage{lipsum}		% Can be removed after putting your text content
\usepackage{graphicx}
\usepackage{setspace}

\usepackage[super,sort&compress,comma]{natbib} 
\usepackage[version=3]{mhchem}
\usepackage{balance}

\title{Microrheology to Probe Smectic Clusters in Bent-core Nematic Liquid Crystals}

\author{ \href{https://orcid.org/0000-0003-4176-2872}{\includegraphics[scale=0.06]{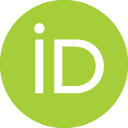}\hspace{1mm}Sathyanarayana Paladugu} \\
	School of Physics, University of Hyderabad, Hyderabad, India.\\
	Department of Physics, Indian Institute of Science Education and Research (IISER) Tirupati, Tirupati, India.\\
	\texttt{sathyapaladugu@gmail.com} \\
	\And
	\href{https://orcid.org/0000-0000-0000-0000}{\includegraphics[scale=0.06]{orcid.png}\hspace{1mm}Supreet Kaur} \\
	Department of Chemical Sciences, Indian Institute of Science Education and Research (IISER) Mohali, India.\\
	\And
	\href{https://orcid.org/0000-0003-3209-3611}{\includegraphics[scale=0.06]{orcid.png}\hspace{1mm}Golam Mohiuddin} \\
	Department of Chemical Sciences, Indian Institute of Science Education and Research (IISER) Mohali, India.\\	%%
	\And
\href{https://orcid.org/0000-0001-6905-0434}{\includegraphics[scale=0.06]{orcid.png}\hspace{1mm}Ravi Kumar Pujala} \\
	Department of Physics, Indian Institute of Science Education and Research (IISER) Tirupati, Tirupati, India.\\
	\And
	\href{https://orcid.org/0000-0000-0000-0000}{\includegraphics[scale=0.06]{orcid.png}\hspace{1mm}Santanu Kumar Pal} \\
	Department of Chemical Sciences, Indian Institute of Science Education and Research (IISER) Mohali, India.\\
	 \And
 \href{https://orcid.org/0000-0003-3144-0300}{\includegraphics[scale=0.06]{orcid.png}\hspace{1mm}Surajit Dhara} \\
	School of Physics, University of Hyderabad, Hyderabad, India.\\
	\texttt{sdsp@uohyd.ernet.in} \\
}

\date{}

\begin{document}

\twocolumn[
  \begin{@twocolumnfalse}

\maketitle

\begin{abstract}
Many bent-core nematic liquid crystals exhibit unusual physical properties due to the presence of smectic clusters, known as ``cybotactic'' clusters in the nematic phase. Effect of these clusters on complex shear modulus ($G^*(\omega)$) of such liquid crystals hitherto unexplored. Here, we study flow viscosities and complex shear modulus of two asymmetric bent-core liquid crystals using microrheology technique. The results are corroborated with the measurements of curvature elastic constants. Compound with shorter hydrocarbon chain (\ce{8OCH3}) exhibit only nematic (N) phase whereas the compound with longer chain (\ce{16OCH3}) exhibits both nematic (N) and smectic-A (SmA) phases.  Our results show that the directional shear modulus of  \ce{16OCH3}, just above the SmA to N transition temperature is strikingly different than \ce{8OCH3}, owing to these smectic clusters. Thus, microrheology enables us to probe smectic clusters in bent-core nematic liquid crystals.
\end{abstract}

 \end{@twocolumnfalse} \vspace{0.6cm}
]

\onehalfspacing

% keywords can be removed

\section{Introduction}
%In a calamitic nematic liquid crystal, the molecules (rod-like) orient along a particular direction called the director $\hat{\mathbf{n}}$~\cite{degennes,oleg}. 
%Physical properties of liquid crystals greatly depend upon the shape of the constituting molecules. 
The bent-core (BC) liquid crystals captivated the interest of many scientists as they continue to exhibit new and interesting physical properties compared to the conventional calamitic liquid crystals made of rod-like molecules~\cite{Takezoe2006,Antal}. For example, bent-core  nematic liquid crystals exhibit negative bend-splay elastic anisotropy ~\cite{sathya1,sathya5,Gortz,Tadapatri,Majumdar,Kaur,Avci} \emph{i.e.} $\delta K=K_{33}-K_{11}<0$, giant flexoelectricity~\cite{Harden,sathya6,Le}, large rotational viscosity~\cite{sathya2,sathya4,sathya5,Dorjgotov}, and induced~\cite{Olivares}, spontaneous biaxiality~\cite{Acharya,Madsen,Jang,Xiang,Lehmann,Yoon,Sluckin}. Since the free rotation of the  molecules around the bow axis is hindered, the BC molecules have strong tendency to form nano-sized smectic clusters, known as cybotactic clusters~\cite{sathya1,sathya3,Tschierske,Dong,Keith,Francescangeli,Domenici}. The unusual properties, some of them reported above have been attributed to the effect of these clusters.  The size of the stable smectic clusters have been measured experimentally using Cryo-transmission electron microscopy (cryo-TEM)~\cite{Zhang}. These smectic clusters are expected to offer both bending modulus and layer compressional modulus that would impact on the complex shear modulus of the nematic phase. However, measurements using conventional  rheological techniques, involving bulk sample would smear out their effect.  Therefore, microrheological techniques, involving a very small amount of sample (typically few microlitres) are important and hitherto unexplored. In this paper, we measure flow viscosities and complex shear modulus and corroborate the results with splay and bend elastic constants of two asymmetric bent-core liquid crystals. Out study shows that microrheological measurements, using self-diffusing microparticles is a sensitive technique to investigate the smectic clusters in bent-core nematic liquid crystals. 

\section{Materials and Methods}
\subsection{Materials}

\begin{figure}[ht]
\includegraphics[width=0.48\textwidth]{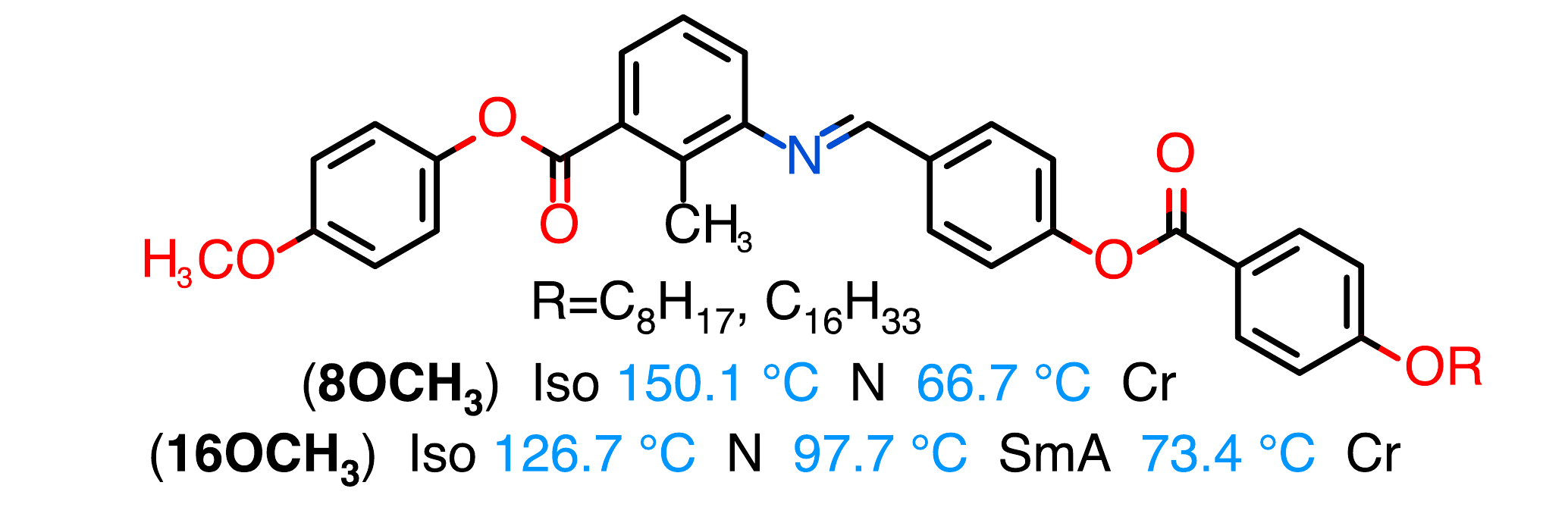}% Here is how to import EPS art
\caption{\label{fig1} Chemical structure of the asymmetric bent-core molecules and phase transition temperatures.}
\end{figure}

Two homologous bent-core compounds, namely \ce{8OCH3}  and \ce{16OCH3} with non-polar (\ce{-OCH3}) moieties were synthesized in our laboratory, where 8 and 16 represents the length of hydrocarbon  chain \ce{R} (see figure~\ref{fig1}). The details of synthesis was reported previously by us~\cite{Kaur1}. The chemical structure of the molecules and the phase transition temperatures of the compounds are shown in figure~\ref{fig1}. They exhibit following phase transitions in cooling: \ce{8OCH3}: I 150.1 $^\circ$C N  66.7 $^\circ$C Cr and \ce{16OCH3}: I 126.7 $^\circ$C N 97.7 $^\circ$C SmA 73.4 $^\circ$C Cr, respectively.

\subsection{Experimental}

\begin{figure}[ht]
\includegraphics[width=0.48\textwidth]{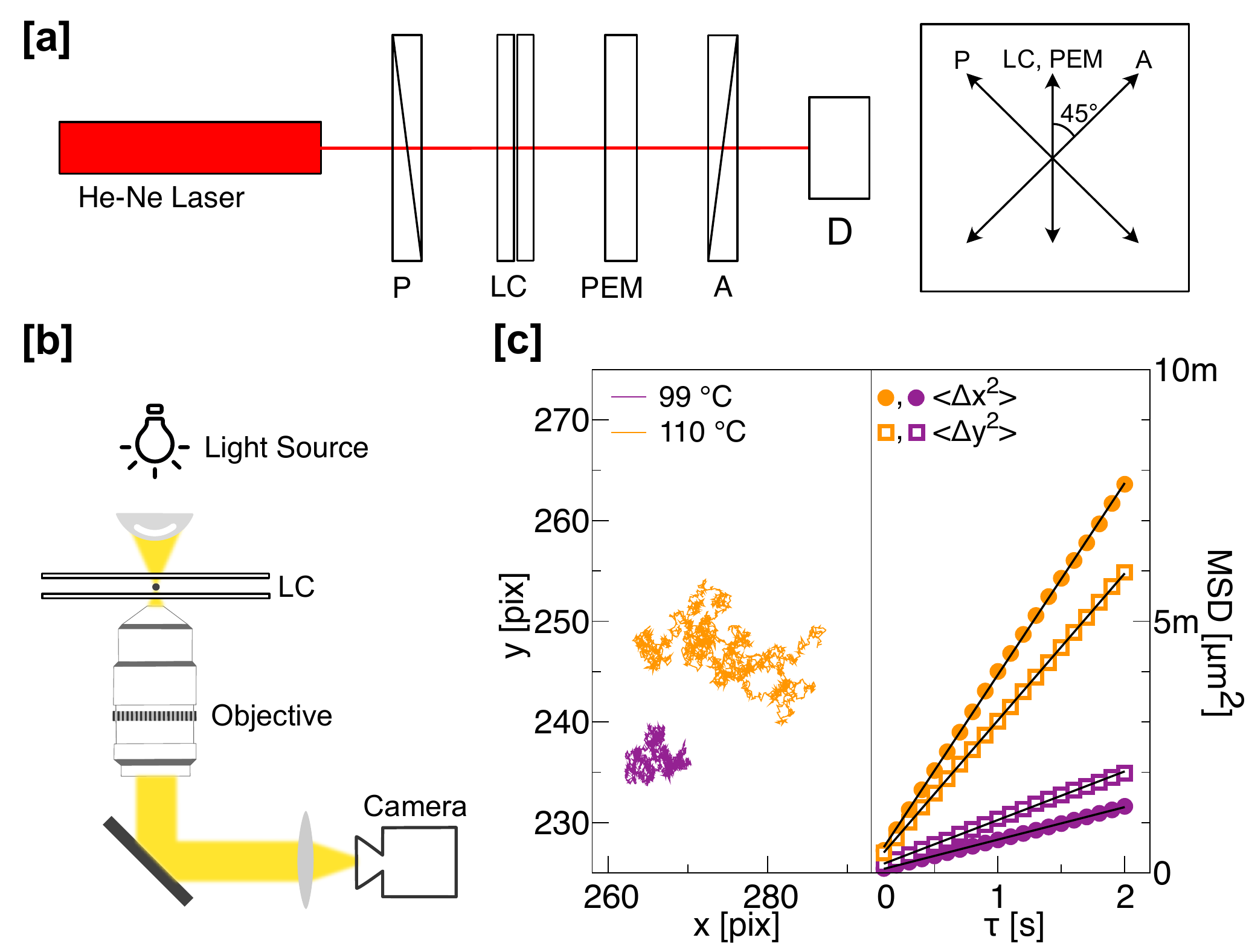}% Here is how to import EPS art
\caption{\label{fig2} [a] Schematic diagram of the experimental setup for measuring birefringence, splay and bend elastic constants. [b] Setup for microrheology studies. [c] Trajectories and time evolution of MSDs of micro-particles at 99 $^\circ$C and 110 $^\circ$C of compound  \ce{16OCH3}. The diffusion coefficients $D_\parallel,~D_\perp$ at 110 $^\circ$C and 99 $^\circ$C (in 10$^\text{-15}~ m^2/s$) are 1.90, 1.46  and 0.32, 0.48 respectively.}
\end{figure}

The experimental cells for aligning the sample were prepared with two patterned indium-tin-oxide (ITO) coated glass plates. The ITO plates were spin coated with polyimide, AL1254 and baked at 180 $^\circ$C for 1 $\mathrm{h}$. AL1254 coated glass plates were then rubbed with a rubbing machine and assembled in antiparallel way to ensure uniform planar alignment. The glass plates were glued together with UV-curable adhesive (NOA65 - Norland Products, Inc) with silica beads as spacers. The thickness of the cells ($d$) were measured with interferometric method using Ocean Optics (USB4000) spectrometer and the typical cell thickness  used in the study was 8~$\mu$m.

The dielectric constant, as a function of voltage ($0.02-20~\mathrm{V}$) at different temperatures in the nematic phase, was measured at 4.111 kHz with Agilent 4980A LCR meter. The perpendicular component of dielectric constant ($\epsilon_\perp$) was obtained from the measurement below the Freedericksz threshold voltage ($\mathrm{V}_\mathrm{th}$) and the parallel component ($\epsilon_\parallel$) was obtained by extrapolating the voltage dependent dielectric constant to $\mathrm{V}\rightarrow\infty$, \emph{i.e.}, $1/\mathrm{V}\rightarrow0$. Simultaneously, the voltage dependent optical retardation was measured with the help of a home built electrooptic setup, consisting of a photo-elastic modulator (Hinds Instruments) and a lock-in amplifier (SRS-830). The experimental diagram of the setup is shown in figure~\ref{fig2}[a].
From the Freedericksz threshold voltage ($\text{V}_\text{th}$) of the retardation measurement, the splay elastic constant ($K_{11}$) was measured, using the relation, $K_{11}=\epsilon_0\Delta\epsilon(\mathrm{V}_\mathrm{th}/\pi)^2$, where $\Delta\epsilon=\epsilon_\parallel-\epsilon_\perp$ is dielectric anisotropy. The bend elastic constant ($K_{33}$) was obtained by fitting the voltage dependent optical retardation, following the method described in Refs.~\cite{deuling,Gruler,Barbero}. The accuracy in the measurement of elastic constant is within 6\%.

The flow viscosities parallel ($\eta_\parallel $) and perpendicular ($\eta_\perp$) to the director were measured by measuring self-diffusion of a silica particle of diameter  3~$\mu$m~\cite{Loudet, Stark}. Before dispersing in LCs, the particles were treated with octadecyldimethyl (3-trimethoxysilylpropyl) ammonium chloride (DMOAP), which align the director normal to the surface~\cite{Skarabot}.  Normal anchoring induces a point defect and stabilises a dipolar director field surrounding the particles as a result of which the particles are levitated in the bulk due to elastic repulsion from the surface~\cite{Pishnyak}. This is advantageous as the data collection is required for longer duration.  The dynamics of the Brownian motion of an isolated silica particle in LCs was recorded through a 60X water immersion objective using a CCD camera (iDs-UI) connected to an Nikon inverted microscope with a frame rate of 20 fps~\cite{Crocker}. The diagram of the experimental setup is shown in the figure~\ref{fig2}[b]. The trajectories of particles are extracted from recorded sequence of images using Python routine, known as Trackpy~\cite{trackpy}. The mean square displacements (MSD) along parallel and perpendicular to the director were measured. The diffusion coefficients parallel ($\mathrm{D}_\parallel$) and perpendicular ($\mathrm{D}_\perp$) to the director were measured from the respective MSDs as a function of the lag time $\tau$, using the equations $<x(\tau)-x(0)^2>=2D_\parallel\tau$ and $<y(\tau)-y(0)^2>=2D_\perp\tau$.  The corresponding viscosities are calculated, using the Stokes-Einstein equation, $\eta_{\parallel,\perp}=k_BT/6\pi R\mathrm{D}_{\parallel,\perp}$, where $k_B$ is the Boltzmann constant, $T$ is the absolute temperature and $R$ is the radius of the microparticle~\cite{Skarabot}.

The complex viscoelastic shear modulus $G^*(\omega)=G'(\omega)+iG''(\omega)$, where $G'$ is storage modulus and $G''$ is loss modulus in frequency ($\omega$) domain, was calculated using the Generalized Stokes-Einstein Equation (GSER)~\cite{Mason}. The GSER in the Fourier domain can be expressed as:
\begin{equation}
G^*(\omega)=G'(\omega)+iG''(\omega)=\frac{k_BT}{\pi Ri\omega\mathcal{F}\{<\Delta{r}^2(t)\} }
\label{eqn:g}
\end{equation}
 The parallel ($G^*_\parallel(\omega)$) and perpendicular components ($G^*_\perp(\omega)$) of the complex shear modulus $G^{*}(\omega)$ were obtained from the MSD along parallel and perpendicular to the nematic director by the equations; $G^*_\parallel(\omega)=\frac{k_BT}{2\pi Ri\omega\mathcal{F}\{<\Delta{x}^2(t)\} }$ and $G^*_\perp(\omega)=\frac{k_BT}{2\pi Ri\omega\mathcal{F}\{<\Delta{y}^2(t)\}}$, respectively~\cite{He}.

\section{Results and discussion}
\subsection{Splay and bend elastic constants}

\begin{figure}[ht]
\includegraphics[width=0.48\textwidth]{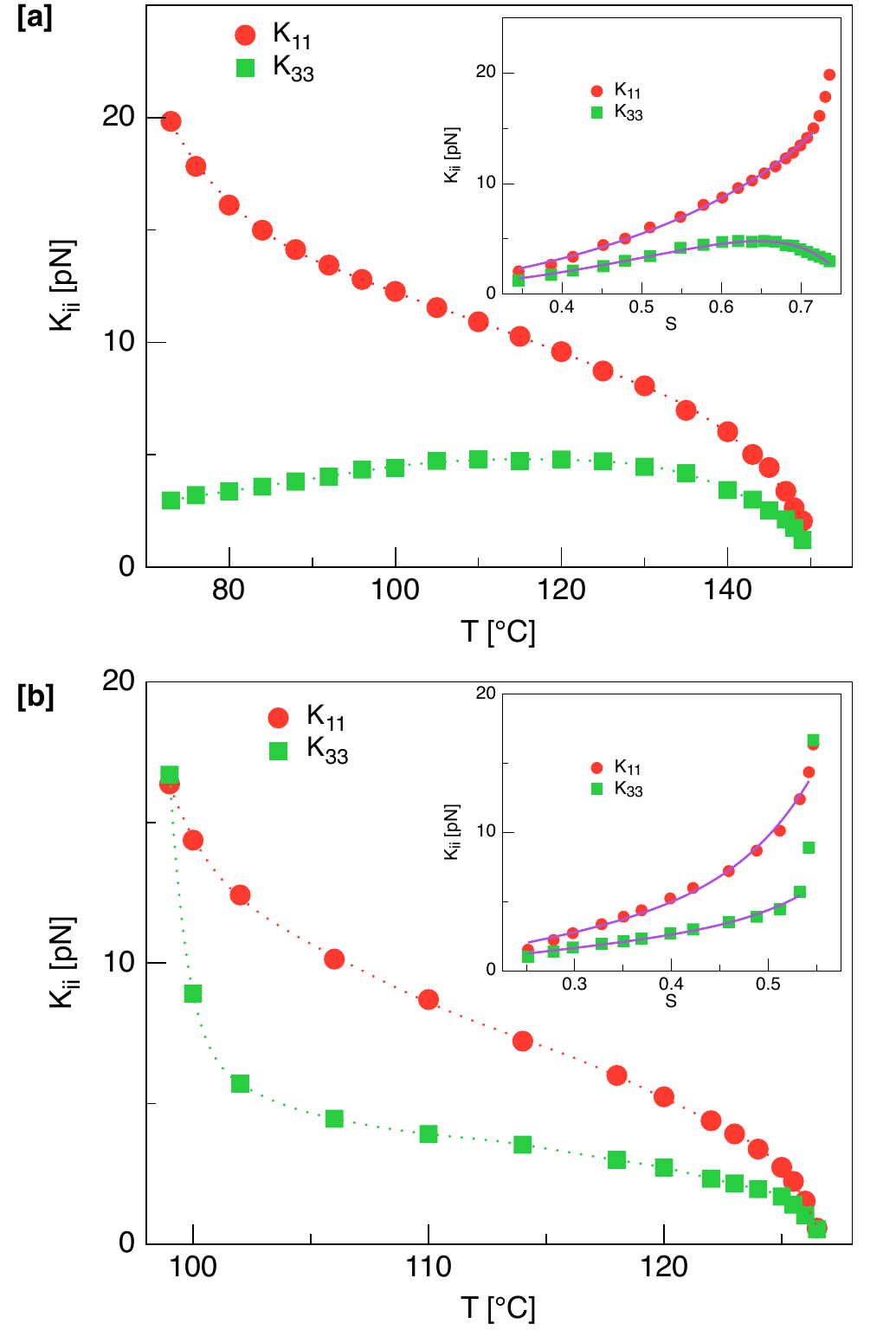}% Here is how to import EPS art
\caption{\label{fig3} [a] Temperature dependent splay ($K_{11}$) and bend ($K_{33}$) elastic constants of \ce{8OCH3}. (inset) $K_{ii}$ as a function of order parameter $S$.  [b] Temperature dependent ($K_{11}$) and ($K_{33}$) of \ce{16OCH3}.(inset) $K_{ii}$ as a function of order parameter $S$. Continuous lines are theoretical fits to equation~\ref{eqn:KvsS}. }
\end{figure}

We begin with the discussion of splay ($K_{11}$) and bend ($K_{33}$) elastic constants of the compounds. Both the compounds exhibit positive birefringence ($\Delta n$) and dielectric anisotropy ($\Delta\epsilon$) (see Supplementary Information). 
 The temperature dependence of $K_{11}$ and $K_{33}$  of both the compounds is shown in figure~\ref{fig3}[a] and 3[b]. 
  The bend-splay anisotropy ($\delta K_{31}=K_{33}-K_{11}$) of the compound \ce{8OCH3} is negative in the entire nematic range and it's magnitude continue to increase with decreasing temperature. Similar negative anisotropy has been reported in many symmetric and asymmetric bent-core nematic liquid crystals ~\cite{kundu,sathya4,Kaur,Avci}. The negative anisotropy is attributed to the coupling of the bent-shape of the molecules with the bend distortion~\cite{sathya1}. For most liquid crystals made of rod-like molecules, $K_{ii}\propto S^2$, where $S$ is the orientational order parameter. To see how $K_{ii}$ of these compounds depend on the order parameter,  we plot $K_{ii}$ as a function of $S$ and show in the inset of figure~\ref{fig3}[a].  The order parameter $S$  was obtained from the temperature dependent birefringence (Supplementary Information). It is noticed that $K_{33}$ shows antagonistic dependence on the order parameter and simply can not be fitted to the second-order on $S$. In particular, $K_{33}$ of \ce{16OCH3} tends to decrease after a pronounced maximum.  The dependence of $K_{ii}$ on $S$ could be better explained using more detailed mean-field calculations given by Berreman and Meiboom~\cite{Berreman}. It introduces a third-order dependence on $S$ and given by:
 %$K_{33}$ increases initially from 0.5 to 4.8 pN and decreases to 3 pN as temperature is lowered, showing a broad maximum at $T-T_{NI}\sim30~^\circ$C. Such kind of behaviour was observed and explained previously~\cite{kundu,sathya4,Kaur,Avci}.
 \begin{equation}
    K_{ii}=K_i^{(2)}S^2+K_i^{(3)}S^3+K_i^{(4)}\frac{S^4}{(1-S)^2}
    \label{eqn:KvsS}
\end{equation}
where $K_i^{n}$, $n=2,3,4$ are fitting parameters. The continuous line in the insets of figure~\ref{fig3}[a] is a theoretical fit to the equation~\ref{eqn:KvsS}. Considering $K^{(2)-(4)}_{i}$'s are temperature independent, the fitting parameters (measured in pN) are found to be $K_1^{(2)}=14.3\pm1.1$, $K_1^{(3)}=13.7\pm1.2$, $K_1^{(4)}=0.7\pm0.1$, and  $K_3^{(2)}=6.4\pm1.1$, $K_3^{(3)}=17.2\pm1.2$, $K_3^{(4)}=-1.8\pm0.1$.
%and $\text{BC}_\text{long}$ molecules are (14.3, 13.7, 0.7), (6.4,17.2,-1.8), (46.5,-67.5,26.3) and (30.9,-48.1,10.8) respectively.

Temperature dependence of $K_{11}$ and $K_{33}$ of the compound \ce{16OCH3}  is shown in figure~\ref{fig3}[b]. In this compound, the bend-splay anisotropy, $\delta K_{31}=K_{33}-K_{11}$ is also negative, except very close to the N-SmA phase transition temperature. In particular, as the N-SmA transition is approached,  $\delta K_{31}$ tends to change sign from negative to positive.
This feature is explained by the strong pre-transitional divergence of $K_{33}$ as bend deformations are incompatible with the equidistance of smectic layers. The dependence of $K_{ii}$ on $S$ is shown in the inset of figure~\ref{fig3}[b]. The excellent fits are obtained with the fit-parameters; $K_1^{(2)}=46.5\pm1.1$, $K_1^{(3)}=-67.5\pm1.2$, $K_1^{(4)}=26.3\pm0.1$ and  $K_3^{(2)}=30.9\pm1.1$, $K_3^{(3)}=-48.1\pm1.2$, $K_3^{(4)}=10.8\pm0.1$).

\subsection{Flow viscosities}
In what follows we measure the flow viscosities of these compounds from the self-diffusion of microparticles following the technique discussed in the experimental section. The measurements are restricted only in the nematic phase as the particles tend to sediment quickly in the isotropic phase. Some representative trajectories of a microparticle and corresponding MSDs at two temperatures, namely 99 $^\circ$C and 110 $^\circ$C for the compound  \ce{16OCH3} are shown in figure~\ref{fig2}[c]. It is found that  at 110 $^\circ$C, $D_{||}/D_{\perp}\simeq1.3$ whereas at 99 $^\circ$C, which is very close to the N-SmA transition temperature, $D_{||}/D_{\perp}\simeq0.7$.  
\begin{figure}[ht]
\includegraphics[width=0.48\textwidth]{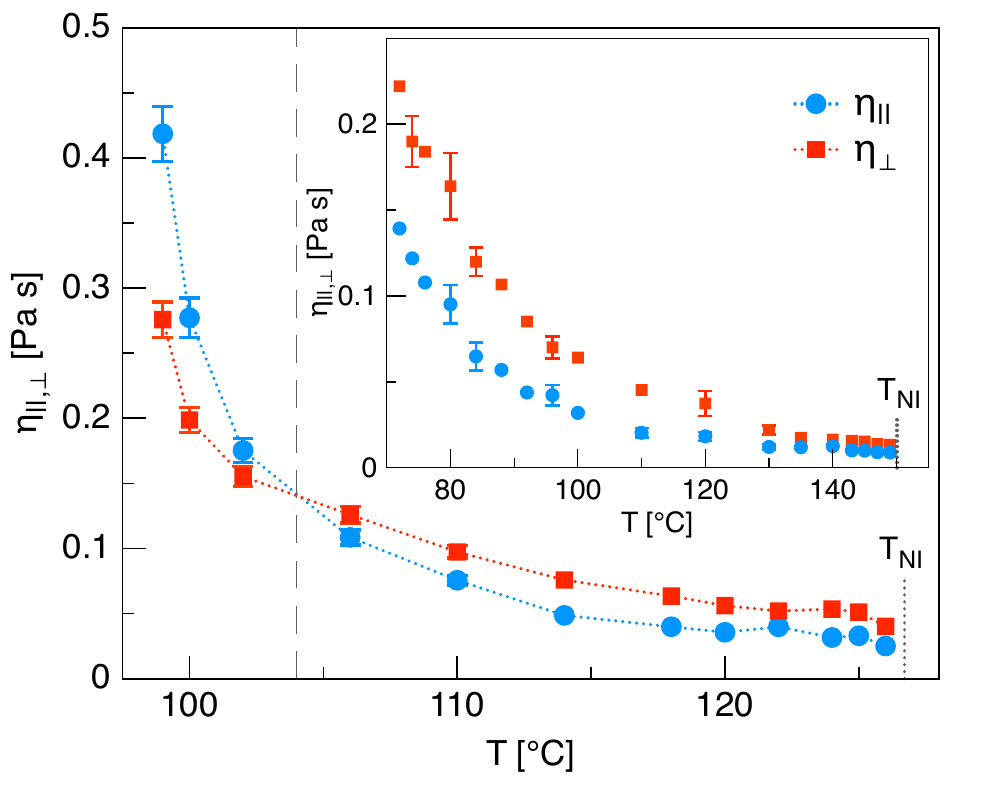}% Here is how to import EPS art
\caption{\label{fig4} Temperature dependent flow viscosities coefficients parallel ($\eta_{||}$) and perpendicular ($\eta_{\perp}$) to the director of \ce{16OCH3}. (inset) Temperature dependent ($\eta_{||}$) and perpendicular ($\eta_{\perp}$) of \ce{8OCH3}. $T_{NI}$ indicates isotropic to nematic phase transition temperature.}
\end{figure}
The temperature dependent viscosities parallel and perpendicular to the director, namely $\eta_\parallel$, $\eta_\perp$ of the  compound \ce{8OCH3}, below the isotropic to nematic transition temperature ($T_{NI}$) are shown in the inset of figure~\ref{fig4}. Both increase with decreasing temperature as expected. It is noticed that in the entire nematic range  $\eta_\perp>\eta_\parallel$, thus viscosity anisotropy ($\delta\eta=\eta_\perp -\eta_\parallel$) is positive. For example, at the lowest temperature ($T=70~^{\circ}$C),  $\delta\eta=\eta_\perp -\eta_\parallel\simeq 100$ mPas.

The temperature dependent viscosities of  compound \ce{16OCH3} in the nematic phase are shown in the figure~\ref{fig4}. 
%The viscosity coefficients $\eta_\parallel\simeq(\eta_2=(\alpha_3+\alpha_4+\alpha_6)/2)$, $\eta_\perp\simeq(\eta_3=\alpha_4)$ takes the values from $\sim25-400$ mPas and increases as the temperature decreases. 
Both $\eta_\parallel$ and $\eta_\perp$ increase with decreasing temperature as expected. Interestingly, it is found that below  $T\approx104~^{\circ}$C, $\delta\eta$ changes sign from positive to negative. For example, at $T=114~^{\circ}$C, $\delta\eta\simeq40$ mPas whereas at $T=98~^{\circ}$C, 
$\delta\eta\simeq-110$ mPas. 
This result can be explained as follows. The compound \ce{16OCH3} has cybotactic clusters in the nematic phase. These clusters are stable compared to the unstable clusters in  liquid crystals with rod-like molecules as the free rotation of the former around the bow axis is reduced sufficiently due to the steric hindrance. Moreover,  the size and volume fraction of these clusters near the N to SmA transition are expected to grow.  The resulting smectic planes of the clusters are orient perpendicular to the director in a planar cell. These  clusters effectively reduces the self-diffusion of the microparticle along the director (perpendicular to the smectic planes) compared to the perpendicular direction (parallel to the smectic planes), resulting negative viscosity anisotropy.
 
 %In fact, the crossover takes place at the temperature, where $K_{33}$ changes curvature and exhibits pretransitional divergence.

\subsection{Complex shear modulus}

\begin{figure}[ht]
 \includegraphics[width=0.45\textwidth]{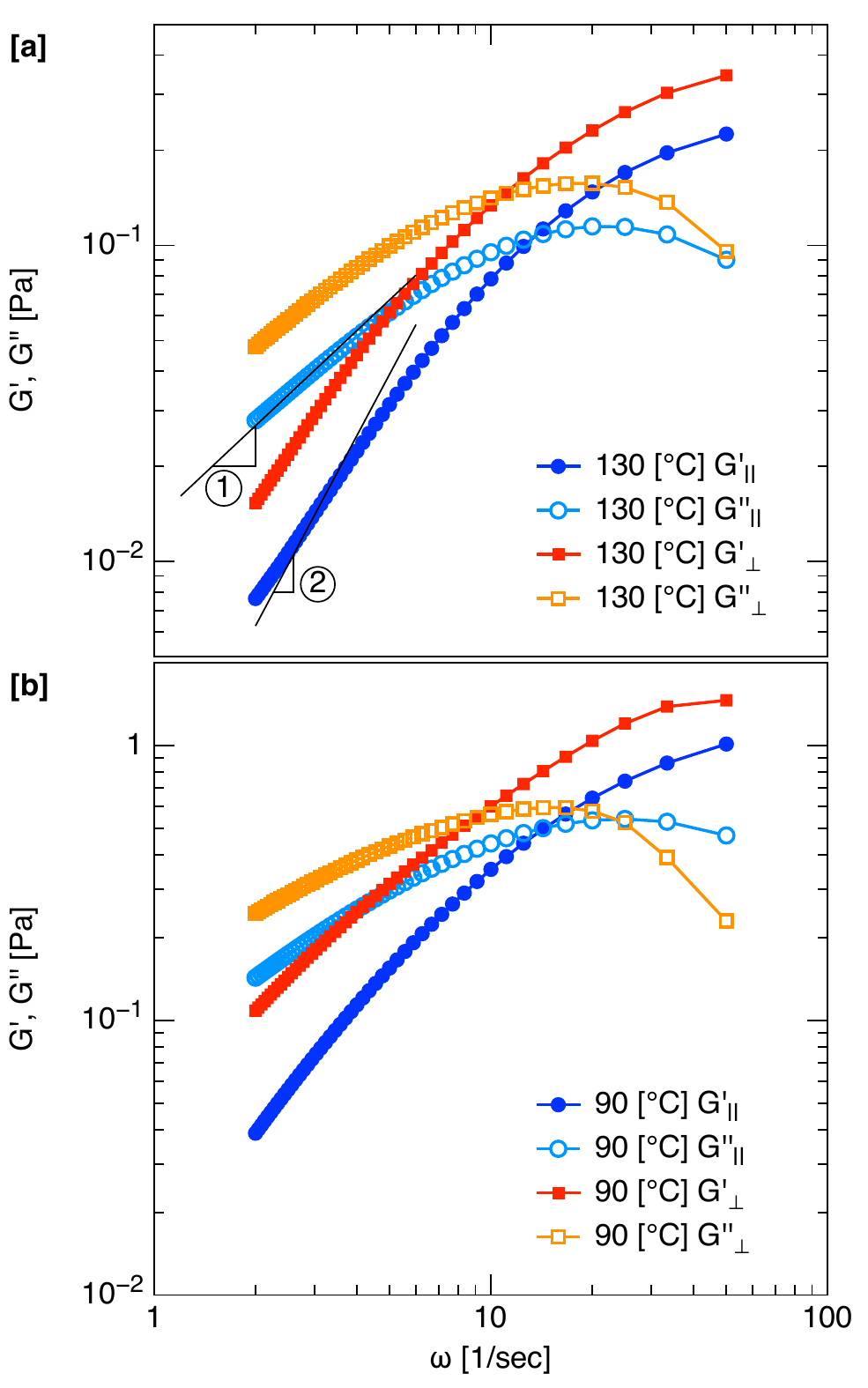}
 \caption{Frequency dependent storage ($G^{'}(\omega)$) and loss modulus ($G^{''}(\omega)$) of \ce{8OCH3}. The subscripts $||$ and $\perp$ denote the components parallel and perpendicular to the director at (a) 130 $^\circ$C and (b) 90 $^\circ$C. {\large \textcircled{\small 1}} and {\large \textcircled{\small 2}} represents the exponents of $\omega$ corresponding to loss and storage moduli, respectively in  Maxwell-fluid model.}
 \label{fig5}
\end{figure}

\begin{figure}[ht]
 \includegraphics[width=0.45\textwidth]{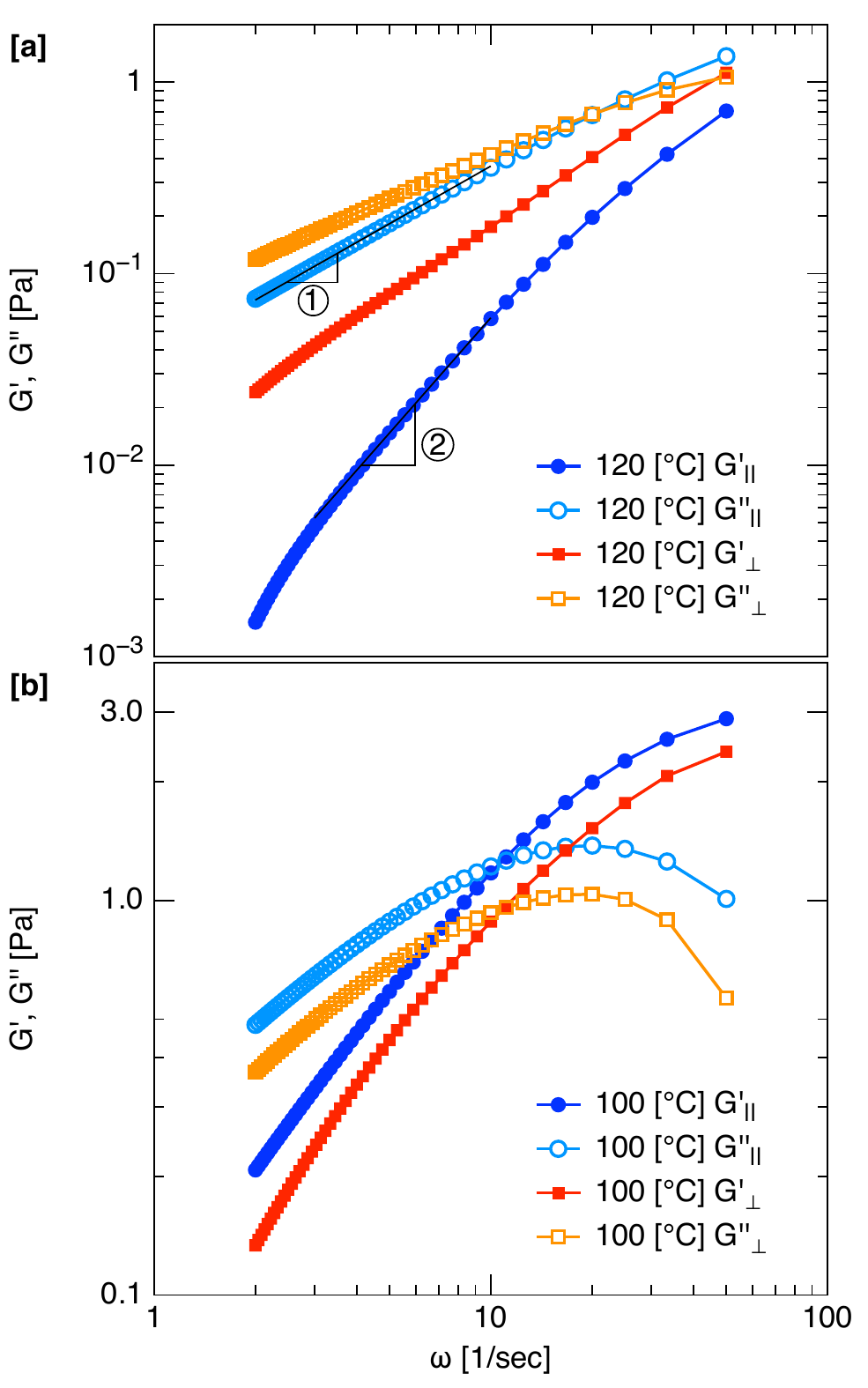}
 \caption{Frequency dependent storage ($G^{'}(\omega)$) and loss modulus ($G^{''}(\omega)$) of \ce{16OCH3}. The subscripts $||$ and $\perp$ denote the components parallel and perpendicular to the director at (a) 120 $^\circ$C and (b) 100 $^\circ$C. {\large \textcircled{\small 1}} and {\large \textcircled{\small 2}} represents the exponents of $\omega$ corresponding to loss and storage moduli, respectively in  Maxwell-fluid model.}
 \label{fig6}
\end{figure}

 The temperature dependent elastic constants and flow viscosities demonstrated the presence of stable cybotactic clusters in the nematic phase of  compound \ce{16OCH3} that contributes greatly to these properties, specially near the N to SmA phase transition. As a next step, we investigate the effect of these clusters on the complex shear modulus $G^{*}(\omega)$. Figure~\ref{fig5} and \ref{fig6} shows the frequency dependence of storage ($G^{'}_{||}(\omega)$ and $G^{'}_{\perp}(\omega)$) and loss modulus ($G^{''}_{||}(\omega)$ and $G^{''}_{\perp}(\omega)$)  along parallel and perpendicular to the nematic director at two different temperatures. Overall, both parallel and perpendicular components exhibit fluid-like behaviour: storage modulus lower than the corresponding loss modulus below a critical frequency. In the low frequency range one observes a behaviour typical of a Maxwell-fluid: $G^{''}_{||,\perp}(\omega)\propto\omega$ and $G^{'}_{||,\perp}(\omega)\propto\omega^2$ (see figure~\ref{fig5}[a] and figure~\ref{fig6}[a]).

 In the nematic phase the perpendicular components are greater than the corresponding parallel components. \emph{i.e.}, $G^{'}_{\perp}(\omega)>G^{'}_{||}(\omega)$ and  $G^{''}_{\perp}(\omega)>G^{''}_{||}(\omega)$ (see figure~\ref{fig5}[a],[b]). Similar behaviour is also seen in compound \ce{16OCH3} at $T=120~^{\circ}$C (figure~\ref{fig6}[a]). However, at lower temperature ($T=100~^{\circ}$C), just above the N-SmA transition temperature, the trend is quite opposite when the relative magnitudes of the parallel and perpendicular components are compared. In particular, $G^{'}_{||}(\omega)>G^{'}_{\perp}(\omega)$ and also $G^{''}_{||}(\omega)>G^{''}_{\perp}(\omega)$ (see figure~\ref{fig6}[b]), which is just opposite to the behaviour seen at higher temperature ($T=120~^{\circ}$C). In addition, the  magnitudes of the directional moduli at $T=100~^{\circ}$C are almost one order of magnitude larger than that measured at $T=120~^{\circ}$C. To bring out the contrasting behaviour we show the ratios $G^{'}_\parallel(\omega)/G^{'}_\perp(\omega)$  and $G^{''}_\parallel(\omega)/G^{''}_\perp(\omega)$ at a fixed $\omega$ in Table-\ref{tbl3:G} and Table-\ref{tbl4:G}. In compound  \ce{8OCH3},  at two different temperatures $G^{'}_\parallel(\omega)/G^{'}_\perp(\omega)<1$  and $G^{''}_\parallel(\omega)/G^{''}_\perp(\omega)<1$ (Table-\ref{tbl3:G}). For compound \ce{16OCH3} at higher temperature ($T=120~^{\circ}$C), both the ratios are also less than 1 (Table-\ref{tbl4:G}). However, at lower temperature ($T=100~^{\circ}$C), $G^{'}_\parallel(\omega)/G^{'}_\perp(\omega)>1$  and $G^{''}_\parallel(\omega)/G^{''}_\perp(\omega)>1$, which is opposite to the feature observed at $T=120~^{\circ}$C.
Thus, parallel components of complex viscoelastic shear modulus of \ce{16OCH3}  exceeds the perpendicular components just above the N-SmA phase transition temperature.

 \begin{table}[ht]
\small
  \begin{tabular*}{0.48\textwidth}{@{\extracolsep{\fill}}lll}
    \toprule
     Temp [$^\circ$C] & $G'_\parallel(\omega)/G'_\perp(\omega)$  & $G''_\parallel(\omega)/G''_\perp(\omega)$  \\
    \hline
     130 & 0.5 & 0.6  \\
     090 & 0.3 & 0.6  \\
    \bottomrule
  \end{tabular*}
  \caption{Ratios of real and imaginary components along parallel and perpendicular to the director of compound \ce{8OCH3} at $\omega$ = 2 sec$^{-1}$.}
  \label{tbl3:G}
\end{table}

\begin{table}[ht]
\small
  \begin{tabular*}{0.48\textwidth}{@{\extracolsep{\fill}}lll}
    \toprule
     Temp [$^\circ$C] & $G'_\parallel(\omega)/G'_\perp(\omega)$ & $G''_\parallel(\omega)/G''_\perp(\omega)$  \\
    \hline
  	120 & 0.1 & 0.6  \\
	100 & 1.5 & 1.3  \\
    \bottomrule
  \end{tabular*}
  \caption{Ratios of real and imaginary components along parallel and perpendicular to the director of compound \ce{16OCH3} at $\omega$ = 2 sec$^{-1}$. }
  \label{tbl4:G}
\end{table}

This distinct feature in the nematic phase can be explained considering the contribution of stable cybotactic clusters.
 In analogy with smectic droplets, the storage modulus of the clusters can be written as  $G^{'}=C\sqrt{(KB/L)}$ where $K$ is the bending modulus, $B$ is the layer compressional modulus,  $L$ is the cluster size and $C$ is a constant~\cite{Panizza,Fujii}. At higher temperature (near NI transition), naturally the clusters are unstable, consequently the contribution due to the layer compressional modulus $B$ is negligible. As the temperature is reduced towards the N-SmA transition, the clusters become stable and their size and the volume fraction increases~\cite{Yuri} as a result of which  $G^{'}$ is enhanced.
Taking $G^{'}=2$ Pa (figure~\ref{fig6}[b]), $K=5$ pN (figure~\ref{fig3}[b]), $C\approx0.2$~\cite{Fujii} and assuming a typical layer compressional modulus $B\approx 10^{6}$ Pa~\cite{Benzekri}, the calculated  cluster size $L\approx 50$ nm, which is very close to the  size measured by Cryo-transmission electron microscopy (cryo-TEM) in similar bent-core liquid crystals~\cite{Zhang}.

\section{Conclusions}
 We studied elastic properties, flow viscosities and directional shear modulus of two bent-core liquid crystals. The bend-splay anisotropy ($\delta K=K_{33}-K_{11}$) of compound \ce{8OCH3} is negative at all temperatures, whereas in \ce{16OCH3}, $\delta K$ tends to change sign just above the N-SmA transition temperature. In compound \ce{8OCH3}, the flow viscosity anisotropy $\delta\eta=(\eta_{\perp}-\eta_{||})>0$, at all temperatures. In case of \ce{16OCH3} it changes sign just above the N-SmA phase transition temperature. The parallel components of complex shear modulus of compound \ce{16OCH3} exceeds the perpendicular components, just above the N-SmA phase transition temperature. This feature is explained considering the contribution of layer compressional modulus of stable smectic clusters in the nematic phase.  Our microrheological studies provided an approximate estimate of smectic clusters size. This technique requires a few microlitres of sample and can be used for probing the effect of nanoscale heterogeneity due to short-range order in liquid crystals and in other soft materials.\\

\balance

\bibliographystyle{unsrt}
\bibliography{article-arxiv}

\begin{thebibliography}{10}

\bibitem{Takezoe2006}
Hideo Takezoe and Yoichi Takanishi.
\newblock Bent-core liquid crystals: Their mysterious and attractive world.
\newblock {\em Japanese Journal of Applied Physics}, 45(2A):597--625, feb 2006.

\bibitem{Antal}
Antal J\'akli, Oleg~D. Lavrentovich, and Jonathan~V. Selinger.
\newblock Physics of liquid crystals of bent-shaped molecules.
\newblock {\em Rev. Mod. Phys.}, 90:045004, Nov 2018.

\bibitem{sathya1}
P.~Sathyanarayana, M.~Mathew, Q.~Li, V.~S.~S. Sastry, B.~Kundu, K.~V. Le,
  H.~Takezoe, and Surajit Dhara.
\newblock Splay bend elasticity of a bent-core nematic liquid crystal.
\newblock {\em Phys. Rev. E}, 81:010702, Jan 2010.

\bibitem{sathya5}
P.~Sathyanarayana, S.~Radhika, B.~K. Sadashiva, and Surajit Dhara.
\newblock Structure–property correlation of a hockey stick-shaped compound
  exhibiting n-sma-smca phase transitions.
\newblock {\em Soft Matter}, 8:2322--2327, 2012.

\bibitem{Gortz}
Verena G{\"{o}}rtz, Christopher Southern, Nicholas~W. Roberts, Helen~F.
  Gleeson, and John~W. Goodby.
\newblock Unusual properties of a bent-core liquid-crystalline fluid.
\newblock {\em Soft Matter}, 5:463--471, 2009.

\bibitem{Tadapatri}
Pramod Tadapatri, Uma~S. Hiremath, C.~V. Yelamaggad, and K.~S. Krishnamurthy.
\newblock Permittivity, conductivity, elasticity, and viscosity measurements in
  the nematic phase of a bent-core liquid crystal.
\newblock {\em The Journal of Physical Chemistry B}, 114(5):1745--1750, 02
  2010.

\bibitem{Majumdar}
M.~Majumdar, P.~Salamon, A.~J\'akli, J.~T. Gleeson, and S.~Sprunt.
\newblock Elastic constants and orientational viscosities of a bent-core
  nematic liquid crystal.
\newblock {\em Phys. Rev. E}, 83:031701, Mar 2011.

\bibitem{Kaur}
S.~Kaur, J.~Addis, C.~Greco, A.~Ferrarini, V.~G\"ortz, J.~W. Goodby, and H.~F.
  Gleeson.
\newblock Understanding the distinctive elastic constants in an oxadiazole
  bent-core nematic liquid crystal.
\newblock {\em Phys. Rev. E}, 86:041703, Oct 2012.

\bibitem{Avci}
Nejmettin Avci, Volodymyr Borshch, Dipika~Debnath Sarkar, Rahul Deb, Gude
  Venkatesh, Taras Turiv, Sergij~V. Shiyanovskii, Nandiraju V.~S. Rao, and
  Oleg~D. Lavrentovich.
\newblock Viscoelasticity{,} dielectric anisotropy{,} and birefringence in the
  nematic phase of three four-ring bent-core liquid crystals with an l-shaped
  molecular frame.
\newblock {\em Soft Matter}, 9:1066--1075, 2013.

\bibitem{Harden}
J.~Harden, B.~Mbanga, N.~\'Eber, K.~Fodor-Csorba, S.~Sprunt, J.~T. Gleeson, and
  A.~J\'akli.
\newblock Giant flexoelectricity of bent-core nematic liquid crystals.
\newblock {\em Phys. Rev. Lett.}, 97:157802, Oct 2006.

\bibitem{sathya6}
P.~Sathyanarayana and Surajit Dhara.
\newblock Antagonistic flexoelectric response in liquid crystal mixtures of
  bent-core and rodlike molecules.
\newblock {\em Phys. Rev. E}, 87:012506, Jan 2013.

\bibitem{Le}
Khoa~Van Le, Fumito Araoka, Katalin Fodor-Csorba, Ken Ishikawa, and Hideo
  Takezoe.
\newblock Flexoelectric effect in a bent-core mesogen.
\newblock {\em Liquid Crystals}, 36(10-11):1119--1124, 2009.

\bibitem{sathya2}
Paladugu Sathyanarayana, Tatipamula~Arun Kumar, Vanka Srinivasa~Suryanarayana
  Sastry, Manoj Mathews, Quan Li, Hideo Takezoe, and Surajit Dhara.
\newblock Rotational viscosity of a bent-core nematic liquid crystal.
\newblock {\em Applied Physics Express}, 3(9):091702, sep 2010.

\bibitem{sathya4}
P.~Sathyanarayana, V.~S.~R. Jampani, M.~Skarabot, I.~Musevic, K.~V. Le,
  H.~Takezoe, and S.~Dhara.
\newblock Viscoelasticity of ambient-temperature nematic binary mixtures of
  bent-core and rodlike molecules.
\newblock {\em Phys. Rev. E}, 85:011702, Jan 2012.

\bibitem{Dorjgotov}
E.~Dorjgotov, K.~Fodor-Csorba, J.~T. Gleeson, S.~Sprunt, and A.~J\'akli.
\newblock Viscosities of a bent‐core nematic liquid crystal.
\newblock {\em Liquid Crystals}, 35(2):149--155, 2008.

\bibitem{Olivares}
J.~A. Olivares, S.~Stojadinovic, T.~Dingemans, S.~Sprunt, and A.~J\'akli.
\newblock Optical studies of the nematic phase of an oxazole-derived bent-core
  liquid crystal.
\newblock {\em Phys. Rev. E}, 68:041704, Oct 2003.

\bibitem{Acharya}
Bharat~R. Acharya, Andrew Primak, and Satyendra Kumar.
\newblock Biaxial nematic phase in bent-core thermotropic mesogens.
\newblock {\em Phys. Rev. Lett.}, 92:145506, Apr 2004.

\bibitem{Madsen}
L.~A. Madsen, T.~J. Dingemans, M.~Nakata, and E.~T. Samulski.
\newblock Thermotropic biaxial nematic liquid crystals.
\newblock {\em Phys. Rev. Lett.}, 92:145505, Apr 2004.

\bibitem{Jang}
Yun Jang, Vitaly~P. Panov, Antoni Kocot, J.~K. Vij, A.~Lehmann, and
  C.~Tschierske.
\newblock Optical confirmation of biaxial nematic (nb) phase in a bent-core
  mesogen.
\newblock {\em Applied Physics Letters}, 95(18):183304, 2009.

\bibitem{Xiang}
Ying Xiang, J.~W. Goodby, V.~G\"{o}rtz, and H.~F. Gleeson.
\newblock Revealing the uniaxial to biaxial nematic liquid crystal phase
  transition via distinctive electroconvection.
\newblock {\em Applied Physics Letters}, 94(19):193507, 2009.

\bibitem{Lehmann}
Matthias Lehmann.
\newblock Biaxial nematics from their prediction to the materials and the
  vicious circle of molecular design.
\newblock {\em Liquid Crystals}, 38(11-12):1389--1405, 2011.

\bibitem{Yoon}
HyungGuen Yoon, Shin-Woong Kang, Matthias Lehmann, Jung~Ok Park, Mohan
  Srinivasarao, and Satyendra Kumar.
\newblock Homogeneous and homeotropic alignment of bent-core uniaxial and
  biaxial nematic liquid crystals.
\newblock {\em Soft Matter}, 7:8770--8775, 2011.

\bibitem{Sluckin}
T.~B.~T. To, T.~J. Sluckin, and G.~R. Luckhurst.
\newblock Biaxiality-induced magnetic field effects in bent-core nematics:
  Molecular-field and landau theory.
\newblock {\em Phys. Rev. E}, 88:062506, Dec 2013.

\bibitem{sathya3}
P.~Sathyanarayana, B.~K. Sadashiva, and Surajit Dhara.
\newblock Splay-bend elasticity and rotational viscosity of liquid crystal
  mixtures of rod-like and bent-core molecules.
\newblock {\em Soft Matter}, 7:8556--8560, 2011.

\bibitem{Tschierske}
Carsten Tschierske and Demetri~J. Photinos.
\newblock Biaxial nematic phases.
\newblock {\em J. Mater. Chem.}, 20:4263--4294, 2010.

\bibitem{Dong}
RONALD~Y. DONG.
\newblock Recent developments in biaxial liquid crystals: An nmr perspective.
\newblock {\em International Journal of Modern Physics B}, 24(24):4641--4682,
  2010.

\bibitem{Keith}
Christina Keith, Anne Lehmann, Ute Baumeister, Marko Prehm, and Carsten
  Tschierske.
\newblock Nematic phases of bent-core mesogens.
\newblock {\em Soft Matter}, 6:1704--1721, 2010.

\bibitem{Francescangeli}
Oriano Francescangeli and Edward~T. Samulski.
\newblock Insights into the cybotactic nematic phase of bent-core molecules.
\newblock {\em Soft Matter}, 6:2413--2420, 2010.

\bibitem{Domenici}
Valentina Domenici.
\newblock Dynamics in the isotropic and nematic phases of bent-core liquid
  crystals: Nmr perspectives.
\newblock {\em Soft Matter}, 7:1589--1598, 2011.

\bibitem{Zhang}
C.~Zhang, M.~Gao, N.~Diorio, W.~Weissflog, U.~Baumeister, S.~Sprunt, J.~T.
  Gleeson, and A.~J\'akli.
\newblock Direct observation of smectic layers in thermotropic liquid crystals.
\newblock {\em Phys. Rev. Lett.}, 109:107802, Sep 2012.

\bibitem{Kaur1}
Supreet Kaur, Golam Mohiuddin, Vidhika Punjani, Raj~Kumar Khan, Sharmistha
  Ghosh, and Santanu~Kumar Pal.
\newblock Structural organization and molecular self-assembly of a new class of
  polar and non-polar four-ring based bent-core molecules.
\newblock {\em Journal of Molecular Liquids}, 295:111687, 2019.

\bibitem{deuling}
Heinz~J. Deuling.
\newblock Deformation of nematic liquid crystals in an electric field.
\newblock {\em Molecular Crystals and Liquid Crystals}, 19(2):123--131, 1972.

\bibitem{Gruler}
Hans Gruler, Terry~J. Scheffer, and Gerhard Meier.
\newblock Elastic constants of nematic liquid crystals.
\newblock {\em Zeitschrift f{\"u}r Naturforschung A}, 27:966,
  2019-12-19T06:04:14.435+01:00 1972.

\bibitem{Barbero}
G~Barbero and L~R Evangelista.
\newblock {\em An Elementary Course on the Continuum Theory for Nematic Liquid
  Crystals}.
\newblock WORLD SCIENTIFIC, 2000.

\bibitem{Loudet}
J.~C. Loudet, P.~Hanusse, and P.~Poulin.
\newblock Stokes drag on a sphere in a nematic liquid crystal.
\newblock {\em Science}, 306(5701):1525--1525, 2004.

\bibitem{Stark}
Holger Stark.
\newblock Physics of colloidal dispersions in nematic liquid crystals.
\newblock {\em Physics Reports}, 351(6):387 -- 474, 2001.

\bibitem{Skarabot}
M.~{\v{S}}karabot and I.~Mu{\v{s}}evi{\v{c}}.
\newblock Direct observation of interaction of nanoparticles in a nematic
  liquid crystal.
\newblock {\em Soft Matter}, 6:5476--5481, 2010.

\bibitem{Pishnyak}
O.~P. Pishnyak, S.~Tang, J.~R. Kelly, S.~V. Shiyanovskii, and O.~D.
  Lavrentovich.
\newblock Levitation, lift, and bidirectional motion of colloidal particles in
  an electrically driven nematic liquid crystal.
\newblock {\em Phys. Rev. Lett.}, 99:127802, Sep 2007.

\bibitem{Crocker}
John~C. Crocker and David~G. Grier.
\newblock Methods of digital video microscopy for colloidal studies.
\newblock {\em Journal of Colloid and Interface Science}, 179(1):298 -- 310,
  1996.

\bibitem{trackpy}
Dan Allan, Casper van~der Wel, Nathan Keim, Thomas~A Caswell, Devin Wieker,
  Ruben Verweij, Chaz Reid, Thierry, Lars Grueter, Kieran Ramos, apiszcz,
  zoeith, Rebecca~W Perry, François Boulogne, Prashant Sinha, pfigliozzi,
  Nicolas Bruot, Leonardo Uieda, Jan Katins, Hadrien Mary, and Aron Ahmadia.
\newblock soft-matter/trackpy: Trackpy v0.4.2, October 2019.

\bibitem{Mason}
T.~G. Mason, K.~Ganesan, J.~H. van Zanten, D.~Wirtz, and S.~C. Kuo.
\newblock Particle tracking microrheology of complex fluids.
\newblock {\em Phys. Rev. Lett.}, 79:3282--3285, Oct 1997.

\bibitem{He}
Jun He, Michael Mak, Yifeng Liu, and Jay~X. Tang.
\newblock Counterion-dependent microrheological properties of $f$-actin
  solutions across the isotropic-nematic phase transition.
\newblock {\em Phys. Rev. E}, 78:011908, Jul 2008.

\bibitem{kundu}
Brindaban Kundu, R.~Pratibha, and N.~V. Madhusudana.
\newblock Anomalous temperature dependence of elastic constants in the nematic
  phase of binary mixtures made of rodlike and bent-core molecules.
\newblock {\em Phys. Rev. Lett.}, 99:247802, Dec 2007.

\bibitem{Berreman}
Dwight~W. Berreman and Saul Meiboom.
\newblock Tensor representation of oseen-frank strain energy in uniaxial
  cholesterics.
\newblock {\em Phys. Rev. A}, 30:1955--1959, Oct 1984.

\bibitem{Panizza}
P.~Panizza, D.~Roux, V.~Vuillaume, C.-Y.~D. Lu, and M.~E. Cates.
\newblock Viscoelasticity of the onion phase.
\newblock {\em Langmuir}, 12(2):248--252, 1996.

\bibitem{Fujii}
S~Fujii, S~Komura, Y~Ishii, and C-Y~D Lu.
\newblock Elasticity of smectic liquid crystals with focal conic domains.
\newblock {\em Journal of Physics: Condensed Matter}, 23(23):235105, may 2011.

\bibitem{Yuri}
Y.~P. Panarin, S.~P. Sreenilayam, J.~K. Vij, A.~Lehmann, and C.~Tschierske.
\newblock Formation and development of nanometer-sized cybotactic clusters in
  bent-core nematic liquid crystalline compounds.
\newblock {\em Beilstein J. Nanotechnol.}, 9(9):1288, 2018.

\bibitem{Benzekri}
M.~Benzekri, T.~Claverie, J.~P. Marcerou, and J.~C. Rouillon.
\newblock Nonvanishing of the layer compressional elastic constant at the
  smetic-a-to-nematic phase transition: A consequence of landau-peierls
  instability?
\newblock {\em Phys. Rev. Lett.}, 68:2480--2483, Apr 1992.

\end{thebibliography}

\end{document}